\documentclass[12pt]{article}
\usepackage{a4,times,amssymb,epsf}

\newcommand{\Ss}[1]{{\sf\small #1}}
\newcommand{\comb}[2]{\left( \! {\,}^{#1}_{#2} \right)}

\begin{document}

\begin{center}
{\sc {\Large {\bf A Combinatorial {\em \sc Bit} Bang Leading to 
Quaternions}}}\footnote{To appear in the Proceedings of the 1997 Helsinki 
Conference on Emergence, Complexity, Hierarchy, Organization (ECHO III).}\\
\vspace{0.2in}
{\small by}\\
{\large Michael Manthey}\\
{\small Department of Mathematics and Computer Science\\
Aalborg University\\
Frederik Bajersvej 7E\\
9220 Aalborg DENMARK\\
manthey@iesd.auc.dk\\
\vspace{0.1in}

\copyright \today}\\

\end{center}

\footnotesize
{\bf Abstract}. This paper describes in detail how (discrete) quaternions - 
ie. the abstract structure of 3-D space - emerge from, first, the Void, and 
thence from primitive combinatorial structures, using only the exclusion and 
co-occurrence of otherwise unspecified events. We show how this computational 
view supplements, and provides an interpretation for, the mathematical 
structures. The build-up is emergently hierarchical, compatible with both 
quantum mechanics and relativity, and can be extended upwards to the 
macroscopic. The mathematics is that of Clifford algebras emplaced in the 
homology-cohomology structure pioneered by Kron. Interestingly, the ideas 
presented here were originally developed by the author to resolve fundamental 
limitations of existing artificial intelligence paradigms.

\normalsize

\section{Introduction}

We find ourselves in a universe of myriad, mystifying, and very nearly 
incomprehensible, complexity. At the same time, contemporary Big Bang 
cosmogenesis tells us that this complexity has apparently {\em emerged} from 
`nothing' - from {\em Void} - via a poorly understood process. In this paper, I 
will attempt to describe a discrete, combinatorial, and computational framework 
for this process. I gladly acknowledge the inspiration of The Combinatorial 
Hierarchy of [Bastin\&Kilmister, Parker-Rhodes], although the material 
presented here differs herefrom in many ways.

How can one get something from nothing? This question is currently being 
framed in terms of the concept of {\em emergence} - that novel properties can 
emerge from simpler constituents, while simultaneously these properties cannot 
be reduced to isolated actions of said constituents. From our point of view, 
the concept of emergence cannot be separated from that of {\em hierarchy}, in 
that emergent properties by definition inhabit a `higher' - that is, more 
complex - level of organization than their constituents. The fact that the 
concept of emergence is controversial is, in our view, a result of the 
residual, but all-pervasive, influence of Newton's physics, which is entirely 
reductionistic.

Nevertheless, a great advantage of the Newtonian view is that it provides an 
intuitive {\em mechanism} for how material entities influence each other: 
momentum exchange, as in the universal image of billiard balls colliding and 
rebounding. The felicity of mechanism is that it provides a blow-by-blow 
description of what is going on, and in so doing fertilizes our imagination 
while simultaneously guiding our modelling choices. The quantum-mechanical 
revolution put an end to this, evaporating `materia' into a cloud of 
probability amplitudes, uncertainty, and non-determinism. Lacking the compass 
of a trustworthy mechanism, we have been collectively doomed to purblindly 
wander the jungles of mathematics, as Einstein well appreciated. For 
mathematics, in spite of appearances, does not really describe {\em how} 
things happen, but rather only their essential {\em what}.

Although we do not argue the case particularly here (see [Manthey94] for some 
initial conjectures), our model implicitly supplies an {\em informational} 
mechanism for quantum mechanics, and at that, one that is compatible with 
relativity theory. The mechanism we present is computational, in the sense that 
it is discrete, combinatorial in character, and described in terms of discrete 
computational operations. However, these operations are not the usual 
arithmetic ``number crunching'' most people associate with computing.
Nor is the computation in question describable in terms of a Turing machine, 
which is - as [Penrose] essentially argued - equivalent to Newtonian 
mechanics. Rather, the concept of an evolving and expanding universe demands a 
{\em distributed multi-process} view. Thus the critical mechanisms are those 
that express the {\em synchronization} between the events constituting the 
various processes. The model that is built up on this basis, and explained in 
the following, we have dubbed the {\em phase web paradigm}; a corresponding 
program, called {\em Topsy}, has been implemented and is available to 
interested persons [www].

A major contribution of this paper is therefore that it shows how to unite 
computational, {\em informational} mechanisms - with the aforementioned 
advantages hereof - with `classical' vector algebra, with surprising and 
intriguing results. In addition, the hierarchical aspect of the basic 
mechanisms shows how it is possible, at least in principle, to tell a detailed 
and rigorous story about the ascent from the microscopic to the macroscopic 
world. 

The outline of the paper is as follows: the next section sketches our 
conceptual, and decidedly computational, framework, its mapping to Clifford 
algebras and a novel hierarchical structure that naturally captures emergent 
phenomena. The following section connects this with our implementation, 
revealing an important ambiguity in the mathematical description versus 
concrete {\em informational} mechanism, which ambiguity is then resolved. We 
then present a combinatorial ``bit bang'' based on the mechanisms introduced, 
and show how quaternions appear.

\newpage

\section{Mechanism, Clifford algebra, and Hierarchy}

Initially, it is crucial to establish the validity of the concept of emergence 
in a mechanistic context.\footnote{Since we are dealing with informational 
mechanism, we prefer the term {\em neo-}mechanistic.} We will see that the 
presence of multiple processes is critical for this purpose.

{\em {\bf The coin demonstration} - Act I. A man stands in front of you with 
both hands behind his back, whilst you have one hand extended in front of you, 
palm up. You see the man move one hand from behind his back and place a coin on 
your palm. He then removes the coin with his hand and moves it back behind his 
back. After a brief pause, he again moves his hand from behind his back, places 
what appears to be an identical coin in your palm, and removes it again in the 
same way. He then asks you, ``How many coins do I have?''.}

It is important at the outset to understand that the coins are {\em formally} 
identical: indistinguishable in every respect. If you are unhappy with this, 
replace them with electrons or geometric points. Also, there are no `tricks' in 
the prose formulation. What is at issue is the fact of 
indistinguishability, and we are simply trying to pose a very simple situation 
where it is indistinguishability, and nothing else, that is in focus.

The indistinguishability of the coins now agreed, the most inclusive answer to 
the question is ``One or more than one'', an answer that exhausts the universe 
of possibilities given what you have seen, namely {\em at least} one coin. 
There being exactly two possibilities, the outcome can be encoded in one bit of 
information. Put slightly differently, when you learn the answer to the 
question, you will per force have received one bit of information.

{\em {\bf The coin demonstration} - Act II. The man now extends his hand and 
you see that there are two coins in it. [The coins are of course identical.]}

You now know that there are two coins, that is, {\em you have received one bit 
of information.} We have now arrived at the final act in our little drama.

{\em {\bf The coin demonstration} - Act III. The man now asks, ``Where did 
that bit of information come from??''}

Indeed, where {\em did} it come from?! Since the coins are indistinguishable, 
seeing them one at a time will never yield an answer to the question. Rather, 
{\em the bit originates in the simultaneous presence of the two coins}. We call 
such a confluence a {\em co-occurrence}.\footnote{At this juncture, we hasten 
to mention that we are dealing here with {\em local} simultaneity, so there is 
no collision with relativity theory. Indeed, Feynman [Feynman65 p.63] argues 
from the basic principle of relativity of motion, and thence Einstein locality, 
that if {\em anything} is conserved, it must be conserved {\em locally.}}

Penrose [Penrose] has argued that computational systems, not least parallel 
ditto, {\em in principle} cannot model quantum mechanics. However, his 
argument is based on Turing's model, which in turn cannot capture 
co-occurrence. 

Notice by the way how the matrix-based formulations of QM neatly get around 
the inherent sequentiality of $y=f(x)$-style (ie. algorithmic) thinking, 
namely by the literal co-occurrence of values in its vectors' and matrices' 
very layouts; and thereafter by how these values are composed {\em 
simultaneously} (conceptually speaking) by matrix operations. Instead of the 
matrix route, we have taken the conceptually compatible one of Clifford 
algebras, which are much more compact, elegant, and general, cf. [Hestenes].

We see from the Coin demonstration that there is information, {\em 
computational information}, available in the universe {\em which {\bf in 
principle} cannot be obtained sequentially.} One can say that the information 
received from observing a co-occurrence is indicative of the fact that two 
states do not mutually exclude each other. 

Co-occurrence and mutual-exclusion are in fact conceptual {\em opposites}, in 
that (say) two events cannot simultaneously both co-occur and mutually 
exclude. The following shows how this insight can be promoted to a concept of 
`action'.

{\em {\bf The block demonstration}. Imagine two `places', $p$ and $q$, each of 
which can contain a single `block'. Each of the places is equipped with a 
sensor, $s_{p}$ respectively $s_{q}$, which can indicate the presence or 
absence of a block.}

The sensors are the {\em only} source of information about the state of their 
respective places and are assumed {\em a priori} to be independent of each 
other, though they may well be correlated. The two states of a given sensor 
$s$ are mutually exclusive, so a place is always either `full', denoted 
(arbitrarily) by $s$, or `empty', denoted by $\tilde{s}$; clearly, 
$\tilde{\tilde{s}}=s$.\footnote{We are working in $\Bbb{Z}_{3}= 
\{0,1,-1=\tilde{1}\}$ rather than the traditional $\Bbb{Z}_{2}=\{0,1\}$. We 
use the visual convention that a sensor written without a tilde is taken to be 
bound to the value $1$, and vice versa; clearly, $\tilde{0}=0$.}

{\em Suppose there is a block on $p$ and none on $q$. This will allow us to 
observe the co-occurrence $s_{p} + \tilde{s}_{q}$. From this we learn that 
having a block on $p$ does not exclude not having a block on $q$. Suppose at 
some other instant (either before or after the preceding) we observe the 
opposite, namely $\tilde{s}_{p} + s_{q}$. We now learn that not having a 
block on $p$ does not exclude having a block on $q$. What can we conclude?}

First, it is important to realize that although the story is built around the 
co-occurrences $s_{p} + \tilde{s}_{q}$ and $\tilde{s}_{p} + s_{q}$, 
everything we say below applies equally to the `dual' pair of co-occurrences 
$s_{p} + s_{q}$ and $\tilde{s}_{p} + \tilde{s}_{q}$.  After all, the 
designation of one of a sensor's two values as `$\sim$' is entirely arbitrary. 
It is also important to realize that the places and blocks are story props: 
all we really have is two two-valued sensors reflecting otherwise unknown 
activities in the surrounding environment. Such sensors constitute the {\em 
boundary} between an entity and its environment in the phase web paradigm.

Returning to the question posed, we know that $s_{p}$ excludes $\tilde{s}_{p}$ 
and similarly $s_{q}$ excludes $\tilde{s}_{q}$. Furthermore, we have observed 
the co-occurrence of $s_{p}$ and $\tilde{s}_{q}$ and vice versa. Since the 
respective parts of one co-occurrence exclude their counterparts in the other 
co-occurrence (cf. first sentence), we can conclude that the co-occurrences 
{\em as wholes} exclude each other.

Take this now a step further. The transition $s_{p} \rightarrow \tilde{s}_{p}$ 
is indicative of some {\em action} in the environment, as is the reverse, 
$\tilde{s}_{p} \rightarrow s_{p}$. The same applies to $s_{q}$. Perceive the 
transitions $s_{p} \leftrightarrow \tilde{s}_{p}$ and $s_{q} \leftrightarrow 
\tilde{s}_{q}$ as two sequential computations, each of whose states consists 
of a single value-alternating bit. By the independence of sensors, 
these two computations are completely independent of each other. At the same 
time, the logic of the preceding paragraph allows us to infer the existence of 
a third computation, a {\em compound} action, with the state transition 
$s_{p} + \tilde{s}_{q} \leftrightarrow \tilde{s}_{p} + s_{q}$, denoted 
$s_{p}s_{q}$. In effect, by combining in this way two single-bit computations 
to yield one two-bit computation, we have lifted our conception of the actions 
performable by the environment to a new, higher, level of abstraction. This 
inference we call {\em co-exclusion}, and can be applied to co-occurrence 
pairs of any arity $>1$ where at least two corresponding components have 
changed.

Notice that the same reasoning applies to the action $s_{p}+s_{q} 
\leftrightarrow \tilde{s}_{p} + \tilde{s}_{q}$, also denoted $s_{p}s_{q}$. The 
two actions are, not surprisingly, {\em dual} to each other, so co-exclusion 
on two sensors can generate two distinct actions. Like co-occurrence, an 
action defined by co-exclusion also possesses an emergent property, generally 
comparable to spin $\frac{1}{2}$ [Manthey94].

Co-exclusion provides a very general mechanism for self-organization: simply 
observe co-excluding co-occurrences, since these then will represent an 
abstraction of the environment. However, the mechanism for actually 
discovering co-exclusions is as yet unspecified. Speaking mechanistically, how 
{\em exactly} does one (eg. the universe) discover the existence of a 
co-exclusive relationship between two co-occurrences? 

Define a co-occurrence in terms of an ``event buffer'' with time-window-size 
$\Delta t$, where true simultaneity requires that $\Delta t = 0$, and larger 
values recognize the factual granularity with one can resolve events and/or 
the time-scale at which an environment varies. Suppose further that event 
identifiers are put into the event buffer as they occur, ie. the new state 
engendered (and labelled) by the action associated with the event is inserted 
into the buffer. Finally, suppose that events in the buffer are successively 
discarded as their residence exceeds $\Delta t$ or the same event-state changes 
again. Clearly, this arrangement guarantees that the state changes contained in 
the buffer all took place within $\Delta t$, and thus  occurred 
`simultaneously' (modulo $\Delta t$). The reader is at this point encouraged to 
ponder the fact that this mechanism in fact solves the problem of discovering 
co-exclusions, and at that, in linear time and space and without 
pre-specification! [Reader pause, for a lovely {\em aha!} experience.]

To see why this claim is true, consider the fact that a sensor's states are 
mutually exclusive, that is, if a sensor is currently in state $s$ then before 
it changed it was in the state $\tilde{s}$. Furthermore, in $\Bbb{Z}_3$ at 
least, the opposite is also true: $\tilde{\tilde{s}}=s$. Hence, since the 
buffer contains the co-occurrence (say) $s_{1}+s_{2}$, and {\em they both just 
changed}, then before they entered the buffer, $\tilde{s}_{1}+\tilde{s}_{2}$ 
obtained. But these two co-occurrences are exactly those necessary to define 
the co-exclusion $s_{1}+s_{2} \longleftrightarrow \tilde{s}_{1}+\tilde{s}_{2}!$ 
The computation time and space are fundamentally linear because they are 
proportional to the buffer size. If we specify that {\em all} events are to 
pass through our event buffer, then the only pre-specification is the arity of 
the co-exclusion. Even this pre-specification can be avoided if all possible 
co-exclusions over the current buffer contents are instantiated as each event 
is entered into the buffer.

\subsection{Co-occurrence and Co-exclusion via Clifford 
Algebras}\label{CliffAlg}

This section presents, very informally, the mathematical foundation of the 
phase web paradigm. The point of departure is to view sensor states as vectors 
instead of scalars, as is conventionally done. 

Let sensor state $s=1$ indicate that sensor $s$ is  currently being 
stimulated, ie. a synchronization token (an informational marker for a state's 
existence) for that state is present, and $s=\tilde{1}$ that $s$ is currently 
{\em not} being stimulated, and hence a token for state $\tilde{s}$ is 
present. Thus the two states of $s$ are represented by their respective 
synchronization tokens, whose respective presences by definition exclude each 
other.

That a set of sensors {\em qua} vectors are orthogonal derives from the fact 
that, in principle, a given sensor says nothing about the state of any other 
sensor. A state of a multi-sensor system is then naturally expressed as the sum 
of the individual sensor vectors, and the state $(s_{a},\tilde{s}_{b}) = 
(1,\tilde{1})$ is written as the vector sum $s_{a}+\tilde{s}_{b}$. Since such 
states represent co-occurrences, it follows that co-occurrences are vector 
sums, usually denoting partial (local) states. Note how the commutativity of 
`$+$' reflects the lack of ordering of the components of a co-occurrence; and 
as well that the co-occurrence $1+\tilde{1}=0$ indicates that the 
interpretation of `zero' is that the components of the sum {\em exclude} each 
other. Because $\Bbb{Z}_{2}$ does not distinguish state-value and exclusion, 
we take our algebra to be over $\Bbb{Z}_{3} = \{0,1,2\}= \{0,1,\tilde{1}\}$.

The next step is to represent {\em actions}. [Manthey94] presents a detailed 
analysis of the group properties of co-occurrences and actions, concluding 
that the appropriate algebraic formalism is a (discrete) Clifford algebra 
[Hestenes], and 
that the state transformation effected by an action is naturally expressed 
using this algebra's vector product. A prime characteristic of this product is 
that it is anti-commutative, that is, for $(s_{1})^2 = (s_{2})^2 = 1$, 
$s_{1}s_{2} = -s_{2}s_{1}$.\footnote{The Clifford product $ab$ can be defined 
as $ab=a\cdot b + a\wedge b$, ie. the sum of the inner ($\cdot$) and outer 
($\wedge$) products, where $a\wedge b = -b\wedge a$ is the oriented area 
spanned by vectors $a,b$. The basis vectors $s_{i}$ of a Clifford algebra may 
have $(s_{i})^2=\pm 1$, and while here we choose $+1$, reasons are appearing 
for choosing $-1$. As long as they all have the same square, it doesn't matter 
for what is said here. Note that $(s_{1}s_{2})^{2}=-1$, so $s_{1}s_{2} \cong 
\surd -1$.} The magnitude of any such product is the area of the parallelogram 
its two components span, and the {\em orientation} of the product is 
perpendicular to the plane of the parallelogram and determined by the ``right 
hand rule''. Applying the Clifford product to a state, one finds - using the 
square-rule and the anti-commutativity of the product given above - that 
\begin{equation}
(s_{1}+s_{2})s_{1}s_{2} = s_{1}s_{1}s_{2} + s_{2}s_{1}s_{2} = s_{2} + 
\tilde{s}_{1}s_{2}s_{2} = \tilde{s}_{1} + s_{2}
\label{(1)}
\end{equation} 
that is, that the result of the {\em action} $s_{1}s_{2}$ is to rotate the 
original state by $90^{o}$, for which reason things like $s_{1}s_{2}$ are 
called {\em spinors}. Thus {\em state change} in the phase web is modelled by 
rotation (and reflection) of the state space, and the effect of an `entire' 
action can be expressed by the inner automorphism $s_{1}s_{2}(s_{1} + 
s_{2})s_{2}s_{1} = \tilde{s}_{1}+\tilde{s}_{2}$, which corresponds to a 
rotation through $180^o$.

One of the felicities of Clifford algebras is that one needn't designate one 
of the axes as `imaginary' and the others as `real'. Rather, the $i$-business 
is implicit  and the  algebra's anti-commutative product neatly bookkeeps the 
desired orthogonality and inversion relationships, no matter how many 
dimensions [ie. sensors (roughly)] are present.

The above 2-spinors are just one example of the vector products available in a 
Clifford algebra - any product of the basis vectors $s_i$ is well-defined, and 
just as $s_{1}s_{2}$ defines an area, $s_{1}s_{2}s_{3}$ defines a volume, etc. 
Being by nature mutually orthogonal, the terms of a Clifford algebra 
\begin{equation}
s_{i} + s_{i}s_{j} +  s_{i}s_{j}s_{k} + \dots + 
s_{i}s_{j}\dots s_{n}
\label{(2)}
\end{equation} 
themselves also define a vector space, which is the space in which we will be 
working (actually, hierarchies of such spaces). [The term (eg.) $s_{i}s_{j}$ 
above, for $n=3$, denotes $s_{1}s_{2}+ s_{2}s_{3}+s_{3}s_{1}$, that is, all 
possible non-redundant combinations.] It is perhaps worth stressing that this 
vector space is the space of the {\em distinctions} expressed by sensors, and 
as such has no direct relationship with ordinary 3+1 dimensional space.

A Clifford product like $s_{1}s_{2}$ reflects both (1) the emergent aspect of a 
phase web action (via its perpendicularity to its components) and (2) its 
ability to act as a meta-sensor (since its orientation is $\pm 1$). Regarding 
(1), the emergence is rooted in the information gleaned from the 
co-occurrences underlying the co-exclusion inference that yields $s_{1}s_{2}$, 
cf. the Coin demonstration. Regarding (2), the co-exclusion inference is an 
{\em abstraction} that produces a single action with two bits of state from two 
lower level actions each possessing a single bit of state. Since this 
abstraction has the same external behavior as its constituent sensors, namely 
$\pm 1$, we can legitimately view it too as a sensor, a {\em meta-}sensor. By 
co-excluding meta-sensors, we can build a new set of abstractions - 
meta-meta-sensors - etc., and thus construct a hierarchy of interwoven 
co-occurrences and exclusions that directly reflects the {\em observed} 
activity of the surrounding environment. This hierarchy is the topic of the 
following.

\subsection{From Clifford Algebra to Hierarchy}

In analogy to $s_{1}, s_{2}$ co-excluding to yield $s_{1}s_{2}$, one might 
expect that the co-exclusion of two meta-sensors, say $s_{i}s_{j}$ and 
$s_{p}s_{q}$, would be modelled by simply multiplying them, to get the 4-action 
$s_{i}s_{j}s_{p}s_{q}$. This turns out however to be inadequate, since although 
by the same logic the co-exclusion of (say) $s_{i}$ and $s_{i}s_{j}$ in a phase 
web expresses explicitly a useful relationship (eg. part-whole), the algebra's 
rules reduce it from $s_{i}s_{i}s_{j}$ to $s_j$, which is simply redundant.

Instead, we take as a clue the fact that {\em change} in a phase web 
occurs via trickling down through the layers of hierarchy, and draw an analogy 
with differentiation. In the present decidedly geometric and discrete context, 
differentiation corresponds to the {\em boundary operator} $\partial$. Define 
$\partial s = 1$ and let $$\partial(s_{1}s_{2}\dots s_{m}) = s_{2}s_{3}\dots 
s_{m} - s_{1}s_{3}\dots s_{m} + s_{1}s_{2}s_{4} \dots s_{m} - \dots 
(-1)^{m+1}s_{1}s_{2}\dots s_{m-1}$$ that is, drop one component at a time, in 
order, and alternate the sign.\footnote{If one takes two components at a time, 
as we will do on occasion later on, then the sign-alternation disappears.} 
Using the algebra's rules as before, one can show that 
$\partial(s_{1}s_{2}\dots s_{m}) = (s_{1}+s_{2} +\dots+ s_{m})s_{1}s_{2}\dots 
s_{m}$ which is exactly the form of equation (1) for what an action does!

Take now equation (2) expressing the vector space of distinctions, segregate 
terms with the same arity, and arrange them as a decreasing series:
\begin{equation}
 s_{i} \stackrel{\partial}{\longleftarrow} s_{i}s_{j} 
\stackrel{\partial}{\longleftarrow} s_{i}s_{j}s_{k} 
\stackrel{\partial}{\longleftarrow} \dots \stackrel{\partial}{\longleftarrow}  
s_{i}s_{j}\dots s_{n-1} \stackrel{\partial}{\longleftarrow} s_{i}s_{j}\dots 
s_{n}
\label{(3)}
\end{equation}
Here as before, $s_{i}s_{j}$ is to be understood as expressing all the 
possible 2-ary forms (etc.), and hence the co-occurrence of pieces of 
similar structure. Each of the individuals is a {\em simplicial complex}, and 
the whole sequence is called a {\em chain complex}, expressing a sequence of 
structures of graded geometrical complexity in which the transition from a 
higher to a lower grade is defined by $\partial$. Furthermore, the entities at 
adjacent levels are related via their group properties - their {\em homology}, 
which we here assume is trivial.

The basic mechanism for expressing change or action in our hierarchical 
context is that of {\em goal-driven} computation. A {\em goal} is a local 
state whose presence causes an action to attempt to change its orientation, 
and a goal will typically be decomposed recursively into subgoals on that 
action's constituents as it trickles down through the $\partial$-hierarchy. 
[Goals differ from the `imperatives' traditionally used in computing - eg. 
\Ss{add x,2} or \Ss{sine(x)} - by not guaranteeing that the indicated 
computation will be achieved, but rather only a `best effort', and success is 
contingent on the state of the environment and the rest of the phase web. 
There is no teleological baggage per se in this concept - `potential' is a 
closer idea.]

It turns out that there is a second structure - a {\em cohomology} - that is 
isomorphic to the homology, but with the difference that arity  {\em 
increases} 
via the $\delta$ (or {\em co-boundary}) operator,\footnote{More precisely, 
$(\sigma_{p},\delta d^{p-1})=(\sigma_{p}\partial,d^{p-1})$, where $\sigma_{p}$ 
is a simplicial complex with arity $p$, and $d^{p}$ the corresponding 
co-complex.} precisely opposite to $\partial$, cf. eqn. (3):
\begin{equation}
s_{i} 
\stackrel{\delta}{\longrightarrow} s_{i}s_{j} 
\stackrel{\delta}{\longrightarrow} s_{i}s_{j}s_{k} 
\stackrel{\delta}{\longrightarrow} \dots \stackrel{\delta}{\longrightarrow}  
s_{i}s_{j}\dots s_{n-1} \stackrel{\delta}{\longrightarrow} s_{i}s_{j}\dots 
s_{n}
\label{(4)}
\end{equation}
Building such increasing complexity is exactly what co-exclusion does. [We 
note that a Clifford algebra satisfies the formal requirements for the 
existence of the associated homology and cohomology.]

It is easily proven that $\partial\partial=0$, and by isomorphism, so also 
$\delta\delta=0$. For example, $\partial\partial(s_{1}s_{2}) = 
\partial(\tilde{s}_{1}+s_{2}) = \tilde{1}+1 = 0$, and similarly, $\partial 
\partial (s_{2}s_{1}) = \partial(s_{1}+\tilde{s}_{2}) = 1+\tilde{1} = 0$. 
Combining these now as the exclusion 
$\partial\partial(s_{1}s_{2}+s_{2}s_{1})$, 
we get $(1+ \tilde{1}) + (\tilde{1}+1) = (1+1)+(\tilde{1}+\tilde{1}) = 0$,
which are the two forms of the input to the determination of a co-exclusion 
relationship. Recalling the event-buffer mechanism for discovering 
co-exclusions, we see, especially if $\Delta t=0$, that this mechanism is a 
realization of the isomorphic $\delta\delta=0$ !

Viewing $\delta$'s abstraction operation informationally, we see that two bits 
($s_{1},s_{2}$) are being encoded in a single bit (the orientation of 
$s_{1}s_{2}$), that is, information is being `abstracted away'. The missing 
bit indicates the {\em phase} of the action, ie. whether the state 
rotation/transformation is $s_{1}+s_{2} \leftrightarrow 
\tilde{s}_{1}+\tilde{s}_{2}$ or $s_{1}+\tilde{s}_{2} \leftrightarrow 
\tilde{s}_{1}+s_{2}$. What will actually occur is however well-defined by the 
other connections $s_{1},s_{2}$ partake in, ie. the boundary conditions of the 
action. Note however that `well-defined' does not necessarily imply 
`deterministic'. Isomorphically, the corresponding $\partial$ operation 
destroys the emergent information in the current state and replaces it by 
non-deterministic outcome.  

\begin{figure}[htbp]
\begin{center}
\leavevmode
\epsfbox{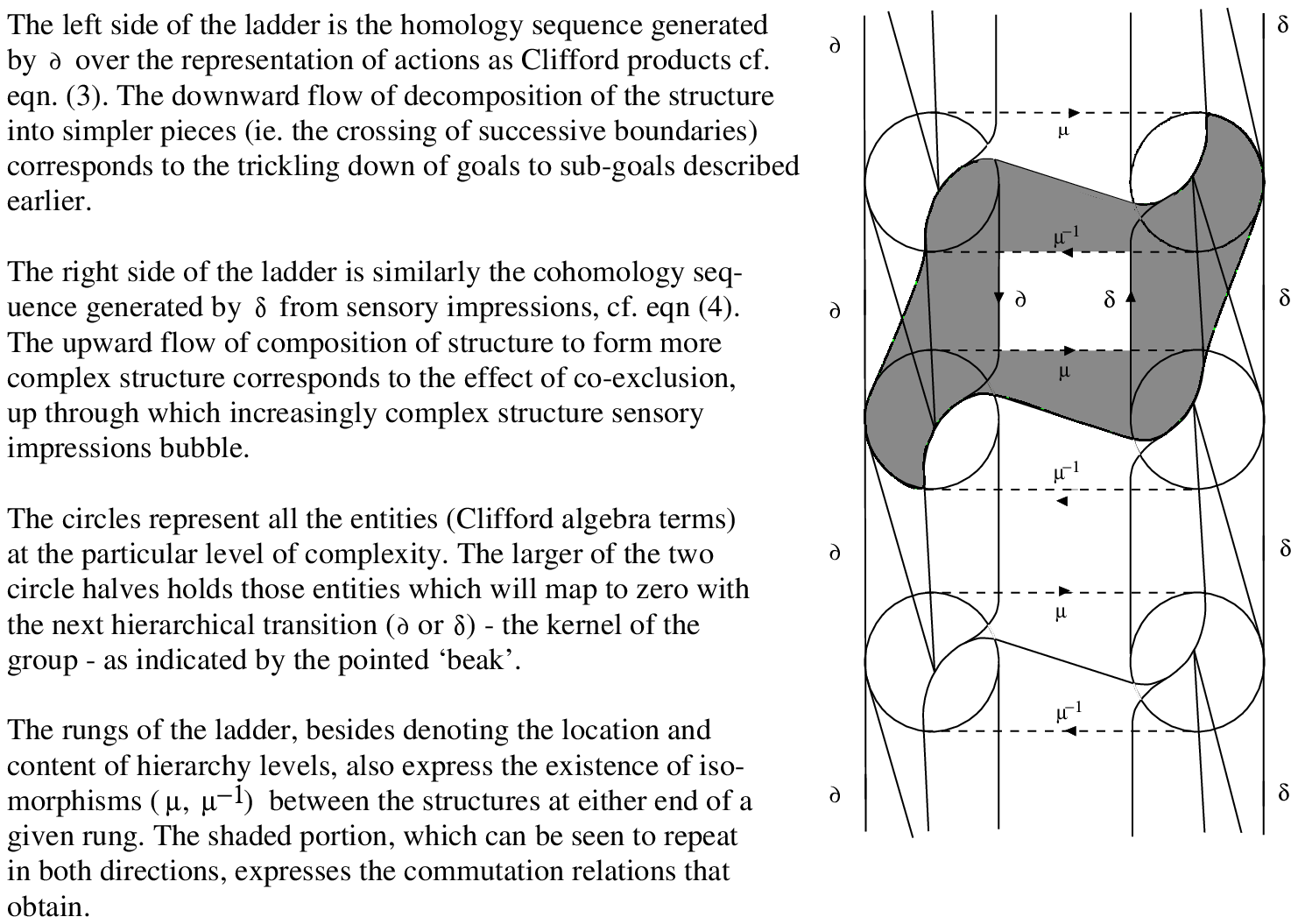}
\caption{Ladder diagram, illustrating homology-cohomology relationships.}
\label{BowdenBase}
\end{center}
\end{figure}

Refer now to Figure \ref{BowdenBase} [Bowden82], which we call a {\em ladder 
diagram}.\footnote{Strictly speaking, $\partial, \delta$, and $\mu/\mu^{-1}$ 
should all be indexed by level: $\partial_{\ell}, \delta_{\ell}, 
\mu_{\ell}/\mu^{-1}_{\ell}$.} 

The shaded shape points out a unique property of the homology-cohomology 
ladder, one that even many topologists seem unaware of, namely that the 
isomorphisms $\mu, \mu^{-1}$ are {\em twisted}, that is, the kernel of the 
group at one end of a rung is mapped by $\mu$ (respectively, $\mu^{-1}$) into 
the non-kernel elements of the group at the other end. [The isomorphisms 
$\mu,\mu^{-1}$ are matrices containing the terms' $\Bbb{Z}_{3}$ coefficients.] 
This property was discovered by [Roth] in his proof of the correctness of 
Gabriel Kron's then controversial methods for analyzing electrical circuits 
[Bowden82], and turns out to have profound implications: the entirety of 
Maxwell's 
equations and their interrelationships can be expressed by a ladder with two 
rungs plus four terminating end-nodes [Bowden], and [Tonti] has - 
independently - shown similar relationships for electromagnetism and 
relativistic gravitational theory. Roth's twisted isomorphism (his term) thus 
reveals the deep structure of the concept of boundary, and shows that the 
complete story requires both homology and cohomology.

\subsection{Generalizing the Twisted Isomorphism Hierarchy}

Each level of a ladder hierarchy, as presented so far, is built entirely from 
entities (ie. sensors) from the level immediately underneath, leading to what 
we call a `pancake' hierarchy. But this is an unnecessary limitation, from 
which we now generalize.

Let $S_{i}$ be the set of sensors at $\delta$-level $i$. Similarly, let 
$G_{i}$ be the set of (sensors expressing the presence of) goals at 
$\partial$-level $i$. A pancake {\em meta} hierarchy of 2-actions can now be 
characterized by $S_{i} = S_{i-1} \times S_{i-1}$, where $\times$ is the 
cartesian product mediated by $\delta$. Other, more general, hierarchical 
forms are now easily seen:

\begin{itemize}
\item $S_{i} = S_{j} \times S_{k}$, $j,k < i$, yielding non-pancake meta 
hierarchies; and of course the product may be over $>$2 levels. Aside from 
this, however, the semantics is roughly as before;

\item $G \times G$, yielding a purely goal-based {\em icarian} hierarchy, 
roughly similar to a function-composition hierarchy;

\item $S \times G$, yielding a combined abstraction over the underlying ladder 
level(s) that we call a {\em morphic} hierarchy.
\end{itemize}

\begin{figure}[htbp]
\begin{center}
\leavevmode
\epsfbox{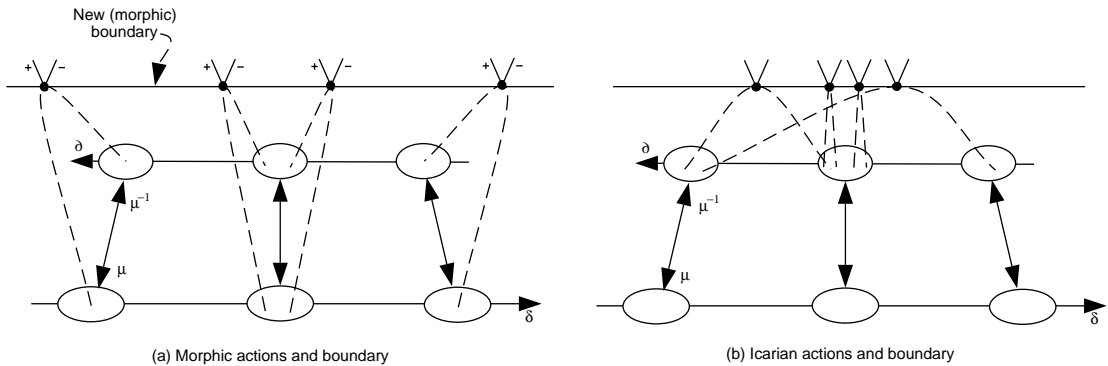}
\caption{Morphic vs. icarian hierarchies.}
\label{MIHier1}
\end{center}
\end{figure}

Figure \ref{MIHier1} illustrates the latter two, and we note that the morphic 
level in (a) may in principle `cross' levels more radically, eg. as (b) does.
We call these generalized forms {\em ortho-hierarchies}.

Icarian actions provide a means for a computation to express, 
self-reflectively, the {\em way} it carries out its goals. Morphic actions 
provide a means for a computation to express, self-reflectively, the 
relationship between $S$ and $G$ that is otherwise buried in $\mu,\mu^{-1}$. 
In the following, we will use the term `meta' to denote all three types of 
abstraction.

\section{Meta-sensor ambiguity and its resolution}

At this point, we have seen two informational mechanisms for emergence - 
co-occurrence and co-exclusion, a mechanism for constructing the latter from 
the former, and a mathematical framework that describes these and their 
composition into a very general hierarchical form. Still missing, though, is 
the mechanism that propagates information up and down this hierarchy.

The key issue here is that the algebraic meta-sensor symbol $s_{1}s_{2}$ codes 
two bits of information - the orientations of $s_{1}$ and $s_{2}$ - into one 
bit - the orientation of $s_{1}s_{2}$. Another way to put this is that the 
algebraic symbol does not distinguish between the duals: $s_{1}+ s_{2} 
\leftrightarrow \tilde{s}_{1}+ \tilde{s}_{2}$ versus $\tilde{s}_{1}+s_{2} 
\leftrightarrow s_{1}+\tilde{s}_{2}$.

Looking more closely at this, there are the following possibilities for a 
state-propagation mechanism. Ie. assuming that we begin in the state 
$s_{1}+s_{2}$, flip the orientation of $s_{1}s_{2}$
\begin{itemize}
\item[1.] When both $s_{1}$ and $s_{2}$ flip.

Here, we encode the two states $s_{1}+s_{2}$ and $\tilde{s}_{1}+\tilde{s}_{2}$ 
as the two possible orientations of $s_{1}s_{2}$. Unfortunately, there is a 
problem: the two dual states  $\tilde{s}_{1}+s_{2}$ and $s_{1}+\tilde{s}_{2}$ 
are not represented at all in the meta-sensor state, and if one of these latter 
states obtains, then the meta-sensor's state is undefined.

If we therefore insist on a distinct meta-sensor for each of the duals, we 
give up any attempt at state compression (abstraction). We also get the problem 
of the very existence of the dual meta-sensor/meta-action...it may not even 
exist yet, and the changing of a single sensor is not enough to trigger its 
discovery (cf. the event window mechanism). This issue revolves around the fact 
that mathematically, all components of the space (here, Clifford algebra terms) 
are always implicitly present when needed, but this is not necessarily the case 
when, as we intend, they directly represent physical entities.

\item[2.] When one of $s_{1}$ or $s_{2}$ flips.

In this encoding, one orientation of $s_{1}s_{2}$ indicates one dual, and the 
other orientation the other dual. However, this does not distinguish between 
the two states within a given dual (Exactly which state are we in? In which 
direction did we rotate?!), and $s_{1}s_{2}$ doesn't flip at all if $s_{1}$ and 
$s_{2}$ flip simultaneously. In effect, we double the rate at which the 
meta-sensor flips to (partially) compensate for the fact that two bits simply 
cannot be encoded in one bit.

This alternative (dubbed `symmetric')succeeds at compressing state only because 
there are exactly 
two duals (which by definition exclude each other).

\item[3.] According to the orientation of $s_{1}$.

Like the preceding alternative, this accepts ambiguity in the phase of the 
current and resulting states, in this case distinguishing one half of the 
$s_{1}\times s_{2}$ plane, but not which quadrant. Mapping $s_{2}$ has a 
conjugal effect. Allowing both maps (to distinct meta-sensors) removes the 
ambiguity at the price of losing abstraction.

\item[$\star$] Note that we have only treated 2-actions. An action of arity $n$ 
possesses $\frac{n}{2}$ duals, so the above issues compound as arity increases. 
We will encounter this in the next section.

\item[$\star$] Note also that the existence of duals creates a {\em naming} 
problem, in that, on the one hand, all the duals of a given action should have 
a common name to reflect this familial relationship, while on the other they 
are all distinct from each other. Thus a given dual's name must be a 2-tuple of 
the form \Ss{(actionId,dualId)}, where \Ss{actionId} is a (commutative) hash of 
(the names of) all the action's constituent sensors (both polarities), whereas 
\Ss{dualId} need only be locally unique.

\end{itemize}

The above list is couched in terms of the bubbling of state change up the 
$\delta$-hierarchy. A similar analysis can be carried out from the point of 
view of goals trickling down the $\partial$-hierarchy, in which case the 
question is: given the goal $s_{1}s_{2} \rightarrow\widetilde{s_{1}s_{2}}$, 
which of the possible subgoals $s_{1} \rightarrow \tilde{s}_{1}$,  $s_{2} 
\rightarrow \tilde{s}_{2}$, or both, should be issued, and when should they be 
retracted?

From either point of view, the second of the above alternatives seems 
preferable, for the following reasons:
\begin{itemize}
\item The phase ambiguity can be seen as a nifty way to model non-deterministic 
outcomes (true also of the third alternative, but not as symmetric).
\item The second alternative satisfies the identity 
$s_{1}s_{2}(s_{1}s_{2}+s_{1}+s_{2})s_{2}s_{1}=s_{1}s_{2}+\tilde{s}_{1}+\tilde{s
}_{2}$, and hence preserves the semantics we arrived at in the preceding 
section.
\item The third alternative has the effect that higher-level abstractions 
continue to ape primitive-level sensors ad infinitum, rather than the new 
exclusions they purport to reflect.
\item The third alternative introduces a new problem: how to choose which of 
the two constituent sensors is to be mapped to the meta-sensor?
\end{itemize}
Unfortunately, the second alternative only works for arity 2, but Nature's 
well-known affinity for symmetry should perhaps not be denied. Therefore we 
accept this hint and will try to solve our problems with arity 2 co-exclusions 
only. Nevertheless, since the choice of propagation model should have 
predictive consequences, this is an issue that can be resolved empirically (and 
we claim a certain amount of empirical support for our choice, as will become 
apparent).

\newpage
\section{The Bit Bang and quaternions}

In this section we present a Big Bang scenario, but where, in contrast to the 
usual version, the expansion is in terms of information. This information is 
the result of making distinctions, and the maker is `the universe' in the guise 
of an initial {\em\bf Void}. In using the latter term, we intend no particular 
a priori interpretation, whether physical, logical, or metaphysical. This said, 
interpreting it in the present context as the vacuum is natural.

Regarding the making of distinctions, and in line with our development thus 
far, we will apply the distinction `co-occur vs. exclude', which pair has the 
distinction-defining property that the two aspects necessarily exclude each 
other. In that the very utterance of one half of a distinction implies the 
other, they become conceptually co-occurrant in yin-yang fashion. This fact is 
in turn captured by the co-exclusion inference, which simultaneously introduces 
the hierarchical moment analyzed in \S 2.3.

Before presenting our {\em Bit Bang}, however, it is appropriate to motivate 
its relevance to our present endeavor. There are two aspects, the first being 
to demonstrate the emergence of space in the form of quaternions. Equally 
important, however, is the more theoretical problem of grounding the endeavor 
as whole. The point here is that in a hierarchical theory, such as the one we 
are presenting, there are two components: the entities that populate the 
various levels (eg. quaternions) and the mechanism of the hierarchy itself, ie. 
the mechanism by which the hierarchy is constructed.

The latter is responsible for the basic properties of the entities, and {\em 
these properties are by definition the same at every level}. We call this 
property of a hierarchical theory {\em level independence}, and it is both the 
blessing and the burden of any hierarchical theory. In the case at hand, the 
basic properties are those of co-occurrences and (co-)exclusions, and 
derivations and implications hereof. Level independence thus implies that the 
phase web's hierarchy should be able to give a reasonable account of its 
creation ab initio, thereby {\em grounding} the entire construction. We 
interpret the ab initio construction as as an informational Big Bang, which 
information is the product of the distinctions afforded by co-occurrence (cf. 
the Coin demonstration) and co-exclusion (cf. the Block demonstration).

On the next page, then, is our Bit Bang. We divide it into a series of steps, 
where each step is intended to follow inevitably from the preceding one. One 
alternative branch is (as noted) from Step 0 to Step 1, where one could 
use $0=\tilde{0}$ instead of $0=0+0$, yielding $\tilde{\mbox{\em\bf 1}}$ at Step 1 
and thereafter reversing the logic of Step 2 to yield {\em\bf 1}; this branch thus 
rejoins the one given at the end of Step 2. In Step 1 one could also ask {\em\bf 
1}$+0$, but this also yields $\tilde{\mbox{\em\bf 1}}$ Finally, Steps 2, 3, and 4 
each `close' a logical level.

Regarding the meta-physical language: what we are trying to convey cries out 
for verbal interpretation, and without such language the mathematical 
expression is more arcane than expressive, expecially in the beginning. 

{\small
\begin{tabular}{lll}
{\em Step} & {\em Symbolically} & {\em Commentary}\\ \hline

0 & {\bf\em Void} = 0 & That about which nothing can be said. Even naming it 
implies\\
		 &&the existence of something that is not {\bf\em Void}. But 
{\bf\em Void} is everything\\
		 &&and nothing, paradoxically simultaneously thinkable and\\
		 &&unthinkable. Physically, {\bf\em Void} is (presumably) the 
vacuum.\\
\ \\
		 &&Mathematically, we might attempt 0 = 0 + 0 = \~0 = \~0 + 
\~0 = 0 + \~0\\
		 &&but even this reifies distinctions we are forbidden: 
`+' implies\\
		 && `parts', and `$\sim$' implies `non-{\bf\em Void}'.\\
\ \\
		 &&But the universe undeniably exists! So from {\bf\em Void} 
there must be\\
		 &&a step. Suppose it was 0 = 0 + 0, ie. the parts are as the 
whole\\
		 &&(starting with $0=\tilde{0}$ yields $\tilde{\mbox{\em\bf 
1}}$, thence {\em\bf 1}). Denote this distinction by\\
		 &&the symbol...\\
\ \\
1 & {\bf\em 1}$_0$    &which means `the same as'. But, having now admitted 
`parts', we\\
		 && must ask, What is {\bf\em 1}$_0$ + {\bf\em 1}$_0$? [We now 
invoke $\Bbb{Z}_{3}$ because: (1) $\Bbb{Z}_{0}$ is\\
		 &&not open to extension, (2) $\Bbb{Z}_{2}$ doesn't distinguish 
{\em\bf Void} from `opposite'\\
		&&so (3) $\Bbb{Z}_{3}$ is the first possibility. 
Since co-occurrence together\\
		&&with exclusion exhaust/fulfill {\em\bf Void} (see 
next step), it appears that $\Bbb{Z}_{3}$\\
		&&is sufficient to all future expansion.] Presuming then 
$\Bbb{Z}_{3}$, the answer is\\
\ \\
2 & $\tilde{\mbox{\em\bf 1}}_0$ &that is, {\bf\em 1}$_0$ is not the same as its 
parts ... $\tilde{\mbox{\em\bf 1}}_0$ means `the opposite of',\\
&&and\\
\ \\ 
  & {\bf\em 1}$_0$ + $\tilde{\mbox{\em\bf 1}}_0$ &means that the parts `the 
same 
as' and `the opposite of' $\equiv$ {\bf\em Void}.\\
 &&Ie. the marriage of sameness and oppositeness exhausts/fulfills\\
&&{\bf\em Void}. Denote this latter distinction, which is {\em new}, by...\\
\ \\
3& {\bf\em 1}$_{1}$ &{\bf\em 1}$_0$ + $\tilde{\mbox{\em\bf 1}}_0 \rightarrow$ 
{\em\bf 1}$_{1}$ is an ``arity 1'' co-exclusion, $\delta_{0}$. [Symbols with 
sub-\\
&&script $= 1$ are in effect `discrete variables in $\Bbb{Z}_{3}$'.]\\
\ \\
 &&Now that we have both {\em sameness} (co-occurrence) and {\em oppositeness} 
\\
 &&(exclusion), we can ask, What is {\bf\em 1}$_0$ + $\tilde{\mbox{\em\bf 
1}}_0$ 
+ {\bf\em 1}$_0$ + $\tilde{\mbox{\em\bf 1}}_0$ = {\bf\em 1}$_1$ + 
$\tilde{\mbox{\em\bf 1}}_1$ ? This\\
 &&distinction is a true (arity $\ge 2$) co-exclusion.\\
\ \\
 &&[This step, and similar ones later, assumes that the {\em\bf Void} 
can/will\\
 &&continue to produce new step-1 instances as needed. Logically, this\\    
 &&is unproblematic; physically, it assumes the same vacuum activity as\\
 &&produced the first instance.]\\
\ \\
 &&The result of the co-exclusion of {\bf\em 1}$_1$ and $\tilde{\mbox{\em\bf 
1}}_1$ is\\
\ \\
4& {\bf\em 1}$_2$ & and, via step 2, $\tilde{\mbox{\em\bf 1}}_2$ follows. Note 
that $(${\em\bf 1}$_{2})^2=(\tilde{\mbox{\em\bf 1}}_1)^{2}=-1$, cf. \S 
\ref{CliffAlg}. \\

\end{tabular}
} 

Clearly, we could continue this listing of distinctions ad infinitum, but we 
choose to end it here, since step 4 has yielded $s_{1}s_{2}$, which is the 
basic quaternion building block.

Via the mappings
\[\mbox{\em\bf 1}_{0}\mapsto 1, \;\;\;\;\tilde{\mbox{\em\bf 1}}_{0} \mapsto 
-1\]
\[ \mbox{\em\bf 1}_{1}\mapsto s, \;\;\;\;\tilde{\mbox{\em\bf 1}}_{1} \mapsto 
\tilde{s}\]
\[\mbox{\em\bf 1}_{2}\mapsto s_{1}s_{2} \mapsto e_{1} \]
\[\mbox{\em\bf 1}_{2}\mapsto s_{2}s_{3} \mapsto e_{2} \]
\[\mbox{\em\bf 1}_{2}\mapsto s_{3}s_{1} \mapsto e_{3} \]
it is easy to verify the defining quaternion relations
\[e_{i}^{2} = -1 \]
\[e_{i}e_{j}=-e_{j}e_{i}, \;\; i \ne j \]
\[e_{1}e_{2}=e_{3}, \;\;e_{2}e_{3}=e_{1},\;\; e_{3}e_{1}=e_{2} \]
whence we have redeemed the promissory note contained in the title of this 
paper.

Notice however that we have only witnessed the emergence of {\em local}  
``3-D-ness''. We are {\em not} claiming (nor do we wish to claim) that this 
3-D-ness is a global Newtonian space with unique origin, {\em nor} even 
relativized multiple ditto. Rather, 3-D-ness is a property of 
co-exclusion-derived objects with sufficient information-carrying capacity 
(``complexity''). The globality of 3-D-ness can only be achieved by the need 
for consistency between objects sharing a given distinction (sensor), and a 
change in the state of a given distinction must therefore propagate through the 
structure. Thus it appears that our construction is entirely consistent with, 
although conceptually `under' or `prior to', general relativity in these 
respects. Moreover, since some distinctions lie below the level at which global 
3-D-ness emerges, changes in these can appear to propagate more rapidly, since 
they are not constrained by the higher level structures (which will 
nevertheless always behave consistently vis à vis such changes). This is the 
phase web's way of reconciling the locality conflict between relativity theory 
and quantum mechanics.

Given that we now have the three quaternion operators and the (local) 3-D 
spatial properties they define, it is natural to ask if the hierarchical 
buildup also can produce the 3-D objects we expect to find in such a space. We 
now address this question.

Our everyday experience tells us that three spatial dimensions can hold 
three-dimensional objects, so we should expect the extension to be 
straightforward. And it is, since $\delta(s_{i}s_{j} + s_{j}s_{k} + s_{k}s_{i}) 
= s_{i}s_{j}s_{k}$ and $s_{i}s_{j}s_{k}$ is an oriented volume (although it 
will develop that this is not {\em quite} right). [We postpone the issue of 
finding some mass to fill this volume.] Notice by the way that this formulation 
requires an arity-3 co-exclusion.

The next issue is how to propagate state up to this new, volumetric, entity. 
The problem is the same as before, except worse: instead of needing to encode 
two bits into one, we now must encode three into one, that is, reflect eight 
possibilites in two. Although we will eventually arrive at a similar solution 
as before (ie. arity 2), the details are instructive.

The table below lists the eight possibilities (viewing the three rightmost 
columns as binary numbers, the first column's numbering is the decimal 
equivalent):
\begin{center}
\begin{tabular}{cccc}
        &$s_{1}s_{2}$&$s_{2}s_{3}$&$s_{3}s_{1}$\\
\ \\
{\bf 7} &    1       &     1      &    1     \\
{\bf 6} &    1       &     1      &$\tilde{1}$\\
{\bf 5} &    1       &$\tilde{1}$ &    1     \\
{\bf 4} &    1       &$\tilde{1}$ &$\tilde{1}$\\
\vspace{1.0mm} \\ \hline
\vspace{1.0mm} \\
{\bf 3} &$\tilde{1}$ &     1      &    1     \\
{\bf 2} &$\tilde{1}$ &     1      &$\tilde{1}$\\
{\bf 1} &$\tilde{1}$ &$\tilde{1}$ &    1     \\ 
{\bf 0} &$\tilde{1}$ &$\tilde{1}$ &$\tilde{1}$\\
\end{tabular}
\end{center}
The pairs {\bf 7,0}, {\bf 6,1}, {\bf 5,2}, {\bf 4,3} are co-exclusions, and are 
distributed symmetrically about the horizontal line. If we simply co-exclude 
these amongst themselves, we will get even more (six, to be exact) so this 
approach diverges. Rather, if we are to use an encoding similar to that of the 
earlier 2-coex case, we must  look at little more closely at the dynamics of 
these entities. One could say that for these four 3-co-exclusions, the dynamics 
is that all three bits flip. What, then, if only one or two flip?

One change at a time yields so-called Grey-coded sequences, and the 
connectivity of the transitions is captured by a unit cube, each of whose 
vertices is labelled by one of the above states. That is, no compression of 
states occurs. We conclude therefore that this kind of distinction is not 
useful (nor is it a co-exclusion, so it's not really a valid distinction 
anyway).

\begin{figure}[htbp]
\begin{center}
\leavevmode
\epsfbox{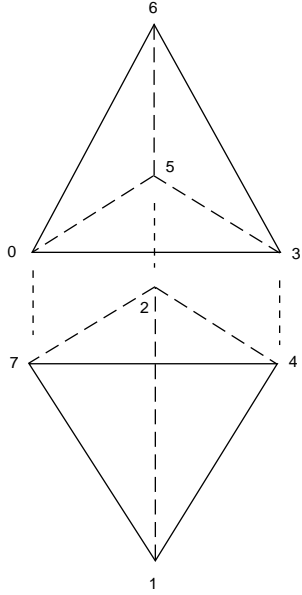}
\caption{The eight states, via 2-flip transitions (denoted by the edges), form 
two tetrahedra of opposite handedness (parity).}
\label{CQtetrahedra}
\end{center}
\end{figure}

In the case where two meta-sensors flip (and one thus remains constant), it 
turns out that there are two disjoint families, {\bf 0, 3, 5, 6} and {\bf 1, 2, 
4, 7}, each defining a tetrahedron, ie. a plane plus a point outside of that 
plane, and hence 3-D orientation. See Figure \ref{CQtetrahedra}. 

Observe now the following:
\begin{itemize}
\item  Half of each family is above/below the line, and the members of the 
halves pair complementarily, so above/below the line each contains all four 
{\em rotation} states around $s_{1}s_{2}$;

\item Each family as well expresses the four possible {\em rotations} around 
$s_{1}s_{2}$, so the same information regarding the two bits that are flipping 
is available both to each family and above/below the line.

\item A transition across the line (ie. between two 3-co-exclusive states) is a 
{\em reflection} (via a factor of $-1$, cf. $ab(a+b)ba=-(a+b)$ );
\end{itemize}

There are three different possible meta-sensors, depending on which of $s_{1}s_{2}, 
s_{2}s_{3}, s_{3}s_{1}$ we choose to be the first column. To make use of our 
symmetric arity-2 meta-sensor construction/restriction, we choose these three 
meta-sensors to be the three co-exclusions (circularly) of $s_{i}s_{j}$ with 
$s_{k}$, ie. $\delta(s_{i}s_{j}+s_{k})$, written $s_{i}s_{j}|s_{k}$. That such a 
meta-sensor form in fact describes $s_{i}s_{j}s_{k}$ is guaranteed by the fact that 
\[\partial(s_{1}s_{2}s_{3}) = -[\partial(s_{1}s_{2})s_{3}+ 
\partial(s_{2}s_{3})s_{1}+ \partial(s_{3}s_{1})s_{2}]\]
where the occurrences of 
$\partial$ on the righthand side can be interpreted as the corresponding events 
(=sensor changes).

Recalling (cf. \S 3) that a symmetric 2-meta-sensor flips when one, but not 
both, of its two constituents flips, the following table shows what happens to 
the three  $s_{i}s_{j}|s_{k}$-meta-sensors when, respectively, one, two, or all 
three constituents flip. {\em Notation}: $\times$ means a flip, a -- means no flip.
\begin{center}
\begin{tabular}{r||ccc|ccc|ccc|c}
What flips& 
$s_{i}s_{j}$&|&$s_{k}$&$s_{j}s_{k}$&|&$s_{i}$&$s_{k}s_{i}$&|&$s_{j}$&Total\\ 
\hline
$s_{i}$&$\times$&&--&--&&$\times$&$\times$&&--&3\\
$s_{i}, s_{j}$&--&&--&$\times$&&$\times$&$\times$&&$\times$&0\\
$s_{i}, s_{j}, s_{k}$&--&&$\times$&--&&$\times$&--&&$\times$&3\\
                     &     &P&       &     &C&       &     &I&&\\
\end{tabular}
\end{center}
\ \\
Since a symmetric meta-sensor only flips when {\em one} of its constituents 
flips, the only changes that count are those that pair a -- with a $\times$, as 
reflected in the rightmost column. Thus a flip of one of the three sensors 
causes all three meta-sensors to flip, whereas a flip of two causes none, and 
when all three base-level constitutents flip, so do all three meta-sensors. We 
now examine each case more closely; the bottom row assigns the names P,C,I 
respectively to the three mixed-level meta-sensors. (It is useful in thinking 
about this to have a picture of a little 3-D coordinate system in mind.)

{\em Only $s_{i}$ flips}. This causes a {\em reflection} of of both the $s_{i}, 
s_{j}, s_{k}$ and PCI coordinate systems.
\\

{\em Both $s_{i}$ and $s_{j}$ flip}. This causes a {\em rotation} in the $s_{i}, 
s_{j}, s_{k}$ coordinate system, but no change in the PCI coordinate system 
(although the changes within P, C, and I are real enough).
\\

{\em All of $s_{i}, s_{j}, s_{k}$ flip}. This causes a {\em reflection} in the both 
the $s_{i}, s_{j}, s_{k}$ and PCI coordinate systems.
\\

Reminding ourselves of the CPT symmetry, this is {\em exactly} what should 
happen had we denoted the P meta-sensor as {\em parity},  the C meta-sensor as 
{\em charge}, and the I meta-sensor as {\em isospin} (which denotes the 
projection of the charge in one of three spatial directions).

In independent support of this identification, we tentatively offer the 
following. There are six quarks, occurring in families of two each, these two 
differing most critically in their charge: $+\frac{2}{3}$ vs. $-\frac{1}{3}$. 
In the mathematical formulation, $s_{1}s_{2}$and $s_{2}s_{3}|s_{3}s_{1}$ are 
operationally equivalent, but in the actual realization, the latter is a 
distinct co-exclusion whose result just happens to have the same effect as 
$s_{1}s_{2}$ but $180^{o}$ out of phase. Because $s_{1}s_{2}$ is half the `size' 
of $s_{2}s_{3}|s_{3}s_{1}$, we assign it charge $-\frac{1}{3}$ and the latter 
$+\frac{2}{3}$. The nicely logical way this works out together with the way 
changes in the three basis sensors $\{s_{1}, s_{2}, s_{3}\}$ are coupled across 
the PCI meta-sensors in CPT-like fashion as just described, argue for 
identifying these three meta-sensors with parity, charge, and 
isospin.\footnote{Having opened Pandora's box here, we speculate that {\em 
mass} is proportional to the number of bits (distinctions) enclosed by a given 
co-exclusion envelope, but a glance at the quarks' empirical values shows that 
there is more to the story.}

We have been tempted to speculate in such matters in order to argue for our 
quaternion construction, and hasten to add that there are many details of the 
above quark structure that must be checked. In any event, it is important for 
the reader to understand that the only degree of freedom in this little game 
lies in {\em how} to arrange the pieces, ie. the number and arity of possible 
co-exclusions and their mappings to corresponding meta-sensor states. {\em All} 
co-exclusions denote distinctions (ie. bits) that the universe can and will 
make, which distinctions people denote by various quantum numbers (generally 
elements of $\Bbb{Z}_3$) eg. spin, parity, and charge. Hence, the combinatorial 
structure {\em must} provide every particle (known or otherwise) with a unique 
and consistent placement in that structure - one misfit means that the whole 
idea dies. In all cases, spin, quaternions and local 3-D-ness, parity, charge, 
and isospin are all clearly seen to be both distributed and emergent, and all 
are properties of the object $s_{1}s_{2}s_{3}$.

\begin{figure}[htbp]
\begin{center}
\leavevmode
\epsfbox{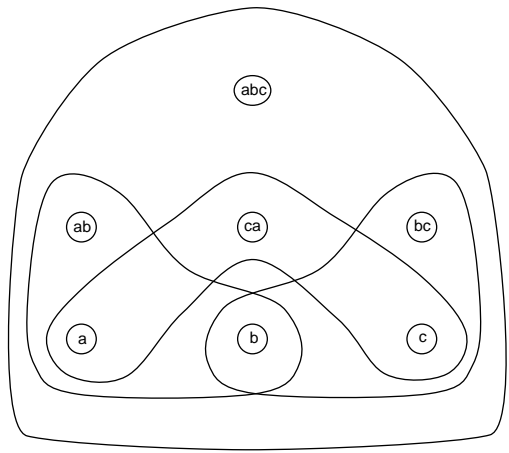}
\caption{Combinatorial Hierarchy categories (dcs's) in $\Bbb{Z}_3$.}
\label{CH_VennDiag}
\end{center}
\end{figure}

There is one final categorization of distinctions we must mention, namely that 
described by The Combinatorial Hierarchy (CH) [Bastin\&Kilmister, 
Parker-Rhodes]. This hierarchy is traditionally constructed in $\Bbb{Z}_2$, but 
there is general agreement that it and the $\Bbb{Z}_3$ Bit Bang presented here 
are in some sense isomorphic. Howsoever, the key point is to examine the number 
of {\em discriminately closed subsets} (dcs's), that is, subsets that close 
under the discrimination operation (in our case, exclusion; in the CH's, 
exclusive-or). These are (cf. Figure \ref{CH_VennDiag})
\[\{s\}\]
\[\{s_{1},s_{2},\{s_{1},s_{2}, s_{1}s_{2}\} \},\]
\[\{s_{1}, s_{2}, s_{3}, \{s_{1},s_{2}, s_{1}s_{2}\},\{s_{2},s_{3}, 
s_{2}s_{3}\}, \{s_{3},s_{1}, s_{3}s_{1}\}, s_{1}s_{2}s_{3}\}\]
\[\dots\]

The key numbers are the successive sums of the dcs cardinalities: $(1), 3, 7, 
127, 2^{127}-1 \longrightarrow 3, 10, 137, 1.7 \times 10^{38}$, ie. column $c$ 
in the table below.
\\

\ \\
\footnotesize
\[
\begin{array}{|c|c|c|c|c|c|}
\hline
 (a) & (b) & (c) & (d) & (e) & (f) \\
\mbox{level} & \mbox{\# symbols} & \mbox{cumulative}&\mbox{map}& \mbox{\# of 
map} & \mbox{comment}\\
 & \mbox{per level} & \mbox{$\sum(b)$}&\mbox{dim.}& \mbox{elements}& \\
\hline
 0  & (1) & (1) & (1) & (1 \times 1) &\\
 1  &  3  &  3  &  4  & 4 \times 4 = 16 & 16 > 7\\
 2  &  7  & 10  &  16 & 16 \times 16 = 256 & 256 > 127\\
 3  & 127 & 137 & 256 & 256 \times 256 = 65536& 65536 < 2^{127}-1\\
 4  &2^{127}-1&2^{127}+136&(256)^{2}&& \mbox{cut-off reached}\\
\hline
\end{array}
\]

\ \\
\normalsize
Column (b) is simply the full number of ways a number of entities (symbols) 
can be combined - 1,2,3... at a time, which is $\sum_{p=1}^{n} \comb{n}{p} = 
2^{n}-1$. This sequence thus counts the number of symbols that can be formed 
from some given set of symbols by aggregation. A second sequence, column 
(d), is related to the number of symbols from column (b) which can via 
discrimination produce the remaining ones at the next level.

Especially the last two numbers in column $c$ are thought provoking: more 
detailed combinatorial calculations yield a corrected value of the inverse fine 
structure constant very near the experimental value (137.0359 674 vs. observed 
137.0359 895(61) ), and similarly for the ratio of the electromagnetic and 
gravitational forces ($2^{127}+136=$ 1.69331$\times 10^{38}$ vs. observed 
1.69358(21)$\times 
10^{38}$), respectively. Interestingly, the sequence in column $b$ cuts off 
after the fourth step, since a symbol-basis of 65536 cannot span a space with 
$2^{127}+136$ elements. See [Noyes] for these and a number of other physical 
constants calculated on this purely combinatorial basis. Note also that the 
dcs's correspond to meta/morphic constructions (cf. \S 2.3) restricted to closure.

\section{Summary and conclusions}

We have described a truly distributed model of computation - the phase web - 
based on the distinction between co-occurrence and mutual exclusion of both 
states and events. This model, by virtue of its acceptance of true concurrency, 
exceeds Turing's model of computation (which conclusion, while not widely 
appreciated, is not controversial in the computer science community). The 
importance of a {\em computational} model, in contrast to so many other kinds, 
is that it provides explicit mechanism, and we argued for the utility of 
mechanism as a tool for reasoning, not least in the context of 20$^{th}$ 
century physical theory. More concretely, without \S 3's search for a mechanism 
for propagating state through the hierarchy, \S 4's quaternion result would 
have been elusive, and perhaps impossible.

The fundamental hierarchy-building operation of co-exclusion - which expresses 
emergent phenomena naturally - turns out to be nicely modelled by Clifford 
algebra's product, which algebra can thereafter be emplaced in the topological 
context of the twisted isomorphism between homology and cohomology. The 
coboundary operator $\delta$ was seen to correspond to, precisely, 
co-exclusion; and the boundary operator $\partial$ to action. Thus the 
hierarchical moment implicit in the co-exclusion operation became 
mathematically explicit.

The fact that each level of the hierarchy is built via the same operation leads 
to the concept of the level independence of phenomena. Level independence is 
what gives power and scope to hierarchical theories, but also carries with it 
the complementary burden of showing that it truly does apply to any level of 
description, or, if you will, empirical fact. Turning this around, {\em if} we 
are to theorize meaningfully about (say) consciousness - which we believe our 
model can accomodate - we should have some reason to believe that our 
theoretical framework is grounded in reality.

To establish this, we modelled the cosmological Big Bang as a process of 
informational expansion deriving from the progressive compounding of 
distinctions, each distinction (co-exclusion) expressing one bit of 
information. Our demonstration in this paper of the emergence from this process 
of local 3-D-ness in the form of quaternions, besides its intrinsic interest, 
thus also allows us to discuss more complex phenomena and systems with rather 
greater confidence.

Relative to the quaternions themselves, we saw that the local 3-D space they 
define requires prior structure possessing sufficient information-carrying 
capacity to express the distinctions associated with 3-D-ness, of which the 
crucial one is parity. We saw that two other distinctions, intertwined with 
parity, appeared at the same time, which, inspired by the CPT theorem, we 
tentatively identified with charge and isospin. This broaching of the topic of 
particle structure gives a another way to test the validity of the points of 
view being advanced in this paper. 

The extension of local 3-D-ness to 3+1 spacetime remains, and the path to be 
followed seems clear, although undoubtedly rocky.

\vspace{3.0mm}
{\em \bf Acknowledgements}.

The basic structure of the Bit Bang was inspired by The Combinatorial Hierarchy 
of [Parker-Rhodes] and [Bastin\&Kilmister], which in turn is based 
on a construction of the integers originally due to Conway. The Coin and Block 
demonstrations are reproduced with IEEE's permission from [Manthey94]. Special 
thanks to Rainer Zimmerman and Achim M{\"u}ller for their gracious hosting of the 
{\em Natura Naturans} `97 workshop in Bielefeldt, where a preliminary version of 
this work was presented.

\newpage
\section*{References}
Bastin, T. and Kilmister, C.W. {\em Combinatorial Physics}. World Scientific, 
1995. ISBN  981-02-2212-2.

Bastin, T. and Kilmister, C.W. ``The Combinatorial Hierarchy and Discrete 
Physics''. Int'l J. of General Systems, special issue on physical theories from 
information (in press).

Bowden, K. ``Physical Computation and Parallelism (Constructive Post-Modern 
Physics)''. Int'l J. of General Systems, special issue on physical theories 
from information (in press).

Bowden, K. {\em Homological Structure of Optimal Systems}. PhD Thesis, 
Department of Control Engineering, Sheffield University UK. 1982.

Feynman, R. {\em The Character of Physical Law}. British Broadcasting Corp. 
1965.

Hestenes, D. and Sobczyk, G. {\em From Clifford Algebra to Geometric Calculus.} 
Reidel, 1989.

Hestenes, D. {\em New Foundations for Classical Mechanics}. Reidel, 1986. The 
first 40 pages contain a very nice, historical introduction to the vector 
concept and Clifford algebras.

Manthey, M. ``Synchronization: The Mechanism of Conservation Laws''. Physics 
Essays (5)2, 1992. 

Manthey, M. ``Toward an Information Mechanics''. Proceedings of the 3rd IEEE 
Workshop on Physics and Computation; D. Matzke, Ed. Dallas, November 1994. ISBN 
0-8186-6715X.

Noyes, H.P. SLAC-PUB-95-7017 and SLAC-PUB-7205,  pp-36-38.

Parker-Rhodes, F. {\em Theory of Indistinguishables - A Search for 
Explanatory Principles Below the Level of Physics}. Reidel. 1981.

Penrose, R. {\em The Emperor's New Mind}. Oxford University Press, 1989. ISBN 
0-19-851973-7.

Roth, J.P. ``An Application of Algebraic Topology to Numerical Analysis: On the 
Existence of a Solution to the Network Problem''. Proc.  US National Academy of 
Science, v.45, 1955.

Tonti, E. ``On the formal structure of the relativistic gravitational theory''. 
Accademia Nazionale Dei Lincei, Rendiconti della classe di Scienze fisiche, 
matematiche e naturali. Serie VIII, vol. LVI, fasc. 2 - Feb. 1974. (In 
english.)

www. Various phase web and Topsy publications, including (soon) code 
distribution, are available via {\em www.cs.auc.dk/topsy}.

\end{document}